\documentclass{appolb}

\begin{document}
\pagestyle{plain}


\newcount\eLiNe\eLiNe=\inputlineno\advance\eLiNe by -1
\title{
TOWARDS DEEP-INELASTIC STRUCTURE \\
FUNCTIONS AT THREE LOOPS}

\author{S. Moch$^a$ and J.A.M. Vermaseren$^b$ 
\address{
\vspace*{-55mm}
\rightline{\normalsize{NIKHEF-2002-006}}
\rightline{\normalsize{TTP-02-19}}
\vspace*{+45mm}
$^a$Institut f{\"u}r Theoretische Teilchenphysik, 
  Universit{\"a}t Karlsruhe \\ 76128 Karlsruhe, Germany \\[1mm]
$^b$NIKHEF Theory Group, 
  Kruislaan 409, 1098 SJ Amsterdam, The Netherlands \\
}}

\maketitle

\begin{abstract}
We report on the status of the calculation of deep-inelastic structure 
functions at three loops in perturbative QCD. 
The method employed allows to calculate the Mellin moments of 
structure functions analytically as a general function of $N$.
As an illustration, we present the leading fermionic contributions 
to the non-singlet anomalous dimension  of $F_2$ at three loops 
and, as a new result, to the non-singlet coefficient function 
of $F_2$ at three loops.
\end{abstract}

\section{Introduction}

Today, structure functions in inclusive deep-inelastic scattering 
are extremely well measured quantities. As a consequence, they 
offer the possibility for very precise determinations of the strong 
coupling $\alpha_s$ and the parton distribution functions.
The high statistical accuracy of the present and upcoming 
experimental measurements demands analyses 
to next-to-next-to leading order (NNLO) of perturbative QCD 
for the structure functions $F_2,F_3$ and $F_L$.

However, the complete NNLO corrections are not fully available yet. 
The two-loop coefficient functions of $F_2,F_3$ and $F_L$ have been 
calculated~\cite{vanNeerven:1991nn,Moch:1999eb}. 
For the three-loop anomalous dimensions only a finite number of fixed 
Mellin moments~\cite{Larin:1997wd}
are presently available. In addition, some information about 
leading fermionic contributions~\cite{Gracey:1994nn,Bennett:1997ch} 
and the small-$x$ limit~\cite{Catani:1994sq} exists.

In the following, we briefly report on the status of the calculation 
of the coefficient functions and the anomalous dimensions to three loops 
in perturbative QCD. Furthermore, we present results for the leading 
fermionic contributions to the non-singlet structure function $F_2$.

\section{Method}

We employ the optical theorem and the operator product expansion (OPE) to 
calculate the deep-inelastic structure functions in Mellin space 
analytically~\cite{Gonzalez-Arroyo:1979df,Kazakov:1988jk} 
as general functions of $N$. 
For the $N$-th Mellin moment of $F_{2}$ we can write 
\begin{eqnarray}
\label{eq:F2mellin}
\displaystyle
F_{2}^N(Q^2)\,=\,     
\int\limits_0^1 dx\, x^{N-2} F_{2}(x,Q^2) \,=\,
\sum\limits_{j=\alpha,{\rm{q, g}}}
C_{2,j}^{N}\left(\frac{Q^2}{\mu^2},\alpha_s\right)
A_{{\rm{P}},N}^j\left(\mu^2\right)\, ,
\end{eqnarray}
where $C_{2,j}^{N}$ denote the coefficient functions and 
$A_{{\rm{P}},N}^j$ the spin averaged hadronic matrix elements 
of singlet operators $O^{\rm q}$, $O^{\rm g}$ and
non-singlet operators $O^{\alpha}$, $\alpha = 1,2,\dots,(n_f^2-1),$ 
of leading twist. 
Both, the coefficient functions and the renormalized operator matrix elements 
in eq.(\ref{eq:F2mellin}) satisfy renormalization group equations 
governed by the same anomalous dimensions $\gamma_{jk}$. 
The anomalous dimensions 
determine the scale evolution of deep-inelastic structure functions.

The calculation of the coefficient functions $C_{2,j}^{N}$ and anomalous
dimensions $\gamma_{jk}$ 
at a given order in perturbation theory amounts to the determination of the $N$-th
moment of all contributing Feynman diagrams with external partons of 
momentum $p$ with $p^2 = 0$ and photons of momentum $q$ with 
$q^2 = -Q^2$.
To achieve this task, we apply the following strategy \cite{Moch:1999eb,Moch:2001fr}.
We set up a hierarchy among all diagrams depending on the number of
$p$-dependent propagators. 
We define basic building blocks (BBB) as diagrams in which the
parton momentum $p$ flows only through a single line in the diagram, 
while composite building blocks (CBB) denote all diagrams with more than 
one $p$-dependent propagator.

Then,  with the help of integration-by-parts
\cite{'tHooft:1972fi} and scaling identities~\cite{Moch:1999eb} 
we determine reduction schemes that map the CBB's of a given topology 
to the BBB's of the same topology or to the CBB's of a simpler topology. 
Subsequently, we use reduction identities that express the BBB's of a 
given topology in terms of simpler topologies.
Working in Mellin space, the reduction equations often involve explicitly 
the parameter $N$ of the Mellin moment. Sometimes, one encounters difference
equations in $N$ for the $N$-th moment $F(N)$ of a diagram,
\begin{eqnarray}
a_0(N) F(N) + a_1(N) F(N-1) + \\ 
\dots + a_n(N) F(N-n) + G(N) &=& 0 \, , \nonumber
  \label{diffeq}
\end{eqnarray}
where $G(N)$ denotes the $N$-th Mellin moment of simpler diagrams. 
First order difference equations can be solved at the cost of one sum 
over $\Gamma$-functions in dimensional regularization. We use 
$D=4-2\epsilon$. The $\Gamma$-functions can be expanded in $\epsilon$
and the sum can be solved to any order in $\epsilon$ in terms of
harmonic sums~\cite{Vermaseren:1998uu,Blumlein:1998if}.
Higher order difference equations could be solved constructively. 
On the mathematical side, the approach to calculate Mellin moments of 
structure functions relies on particular mathematical concepts~\cite{Vermaseren:2000we}, 
such as harmonic sums~\cite{Vermaseren:1998uu,Blumlein:1998if} and 
our ability to set up and solve the difference equations as nested sums 
in $N$. 

\section{Leading fermionic contributions}

To illustrate the method, we discuss the leading fermionic contributions 
to the non-singlet structure function. At three loops, they are proportional
to $n_f^2$, with $n_f$ being the number of massless fermions. 
These contributions form a gauge-invariant subset, but do not involve yet 
any genuine three-loop topologies. Therefore, in the sense of the reduction 
strategy sketched above, the $n_f^2$-terms are easier to calculate.

The result for the $n_f^2$-contribution  
to the non-singlet anomalous dimension $g_{\rm qq}^{(2),\rm ns}$ 
at three loops is known from the work of Gracey \cite{Gracey:1994nn}.
It is given by,
\begin{eqnarray}
\label{eq:gqq2nf}
        g_{\rm qq}^{(2),\rm ns} &=&
         C_F n_f^2  \Biggl(
            {17 \over 9}
          + {32 \over 9} {1 \over N\!+\!1}
          - {88 \over 27} {1 \over (N\!+\!1)^2}
          + {8 \over 9} {1 \over (N\!+\!1)^3}
          - {32 \over 9} {1 \over N}
\nonumber
\\
&&
          + {88 \over 27} {1 \over N^2}
          - {8 \over 9} {1 \over N^3}
          - {16 \over 27} S_{1}(N)
          - {80 \over 27} S_{2}(N)
          + {16 \over 9} S_{3}(N)
           \Biggr)
\, ,
\end{eqnarray}
with $C_F=(N^2-1)/(2N)$, which is $4/3$ for QCD. 

As a new result, we give here the $n_f^2$-contribution to the non-singlet 
coefficient function $c_{2,\rm qq}^{(3),\rm ns}$ at three loops for the
flavour class, where both photons couple to the external quark.
Strictly speaking, the three-loop coefficient functions contribute 
in a perturbative expansion only at next-to-next-to-next-to-leading order (NNNLO). 
However, the result illustrates nicely that our method will not only provide the 
anomalous dimensions, which are proportional to the single
pole in $\epsilon$ in dimensional regularization, but also  
the coefficient functions which are determined by the finite terms,
since our approach is (at least in principle) not limitated to a given order in $\epsilon$.

In terms of harmonic sums up to weight four, 
our result for the $n_f^2$-contribution to $c_{2,\rm qq}^{(3),\rm ns}$ reads
\begin{eqnarray}
\label{eq:c2qq3nf}
\lefteqn{
        c_{2,\rm qq}^{(3),\rm ns} \, =} 
\nonumber
\\
&&
         C_F n_f^2  \Biggl(
          - {9517 \over 486}
          - {8 \over 9} \zeta_3
          + {36748 \over 729} {1 \over N\!+\!1}
          + {16 \over 27} {\zeta_3 \over N\!+\!1} 
          - {4384 \over 81} {1 \over (N\!+\!1)^2}
\nonumber
\\
&&
          + {2360 \over 81} {1 \over (N\!+\!1)^3}
          - {184 \over 27} {1 \over (N\!+\!1)^4}
          + {16 \over 3} {S_{1}(N\!+\!1) \over (N\!+\!1)^3} 
          - {544 \over 27} {S_{1}(N\!+\!1) \over (N\!+\!1)^2} 
\nonumber
\\
&&
          - {32 \over 9} {S_{1,1}(N\!+\!1) \over (N\!+\!1)^2} 
          + {32 \over 9} {S_{2}(N\!+\!1) \over (N\!+\!1)^2} 
          + {2240 \over 81} {S_{1}(N\!+\!1) \over N\!+\!1} 
          + {272 \over 27} {S_{1,1}(N\!+\!1) \over N\!+\!1}
 \nonumber
\\
&&
          + {16 \over 9} {S_{1,1,1}(N\!+\!1) \over N\!+\!1} 
          - {16 \over 9} {S_{1,2}(N\!+\!1) \over N\!+\!1}
          - {272 \over 27} {S_{2}(N\!+\!1) \over N\!+\!1} 
          - {16 \over 9} {S_{2,1}(N\!+\!1) \over N\!+\!1} 
\nonumber
\\
&&
          + {16 \over 9} {S_{3}(N\!+\!1) \over N\!+\!1} 
          - {11170 \over 729} {1 \over N}
          - {16 \over 27} {\zeta_3 \over N}
          + {1204 \over 81} {1 \over N^2}
          - {992 \over 81} {1 \over N^3}
          + {184 \over 27} {1 \over N^4}
\nonumber
\\
&&
          - {16 \over 3} {S_{1}(N) \over N^3} 
          + {232 \over 27} {S_{1}(N) \over N^2} 
          + {32 \over 9} {S_{1,1}(N) \over N^2} 
          - {32 \over 9} {S_{2}(N) \over N^2} 
          - {644 \over 81} {S_{1}(N) \over N} 
\nonumber
\\
&&
          - {104 \over 27} {S_{1,1}(N) \over N} 
          - {16 \over 9} {S_{1,1,1}(N) \over N} 
          + {16 \over 9} {S_{1,2}(N) \over N} 
          + {104 \over 27} {S_{2}(N) \over N} 
\nonumber
\\
&&
          + {16 \over 9} {S_{2,1}(N) \over N} 
          - {16 \over 9} {S_{3}(N) \over N} 
          + {8714 \over 729} S_{1}(N)
          + {32 \over 27} S_{1}(N) \zeta_3
          + {940 \over 81} S_{1,1}(N)
\nonumber
\\
&&
          + {232 \over 27} S_{1,1,1}(N)
          + {32 \over 9} S_{1,1,1,1}(N)
          - {32 \over 9} S_{1,1,2}(N)
          - {232 \over 27} S_{1,2}(N)
\nonumber
\\
&&
          - {32 \over 9} S_{1,2,1}(N)
          + {32 \over 9} S_{1,3}(N)
          - {860 \over 27} S_{2}(N)
          - {536 \over 27} S_{2,1}(N)
          - {64 \over 9} S_{2,1,1}(N)
\nonumber
\\
&&
          + {64 \over 9} S_{2,2}(N)
          + {2440 \over 81} S_{3}(N)
          + {32 \over 3} S_{3,1}(N)
          - {368 \over 27} S_{4}(N)
           \Biggr)\, .
\end{eqnarray}
This result agrees with the one of the fixed Mellin
moment calculation~\cite{Larin:1997wd} for 
$N=2,\dots,14$. It may be used to improve 
approximations~\cite{vanNeerven:1999ca} 
to the full functional form of $c_{2,\rm qq}^{(3),\rm ns}$, although 
the $n_f^2$-terms are numerically not the most dominant contribution.

\section{Conclusions}

The present approach based on the OPE, to calculate the Mellin moments 
of structure functions allows for the calculation of the complete 
coefficient functions and the anomalous dimensions at three loops.

As a next step, one can consider the subleading fermionic contributions 
at three loops proportional to $n_f$. 
The determination of these terms already relies on major parts of the 
complete reduction scheme as it requires the calculation of several 
genuine three-loop topologies of the Benz type as well as 
the calculation of two-loop topologies with a self-energy insertion. 
The results for the $n_f$-contribution to the non-singlet anomalous 
dimension will be presented elsewhere.

\end{document}